\documentclass[journal,10pt]{IEEEtran}

\usepackage{cite}
\usepackage{amsmath,amssymb,amsfonts}
\usepackage{array}
\usepackage{algorithm}
\usepackage{algpseudocode}
\usepackage{graphicx}
\usepackage{textcomp}
\usepackage{xcolor}
\usepackage{booktabs}
\usepackage{url}

\begin{document}

\title{ST-CGMDiff: Sparse-Conditioned Cross-Frequency Spatio-Temporal Latent Diffusion for Dynamic 3D Channel Gain Map Reconstruction}

\author{Tianxiang Li, Yilong Chen, Han Hu, Yong Zeng,~\IEEEmembership{Fellow, IEEE}, and ~\mbox{Jie Xu},~\IEEEmembership{Fellow, IEEE}
\thanks{Tianxiang Li is with Chien-Shiung Wu College, Southeast University, Nanjing 211102, China, and also with the School of Science and Engineering and the Future Network of Intelligence Institute, The Chinese University of Hong Kong (Shenzhen), Shenzhen 518172, China (e-mail: 213233636@seu.edu.cn).}%
\thanks{Yilong Chen and Jie Xu are with the School of Science and Engineering and the Future Network of Intelligence Institute, The Chinese University of Hong Kong (Shenzhen), Shenzhen 518172, China (e-mail: chenyilong@cuhk.edu.cn; xujie@cuhk.edu.cn).}%
\thanks{Han Hu is with the School of Information and Electronics, Beijing Institute of Technology, Beijing 100081, China (e-mail: hhu@bit.edu.cn).}%
\thanks{Yong Zeng is with the National Mobile Communications Research Laboratory, Southeast University, Nanjing 210096, China, and also with Purple Mountain Laboratories, Nanjing 211111, China (e-mail: yong\_zeng@seu.edu.cn).}%
\thanks{Part of this paper has been submitted to IEEE Vehicular Technology Conference (VTC) 2027-Spring~\cite{bPreliminary}.}%
\thanks{Corresponding authors: Yilong Chen and Jie Xu.}%
}

\maketitle

\begin{abstract}
Accurate and efficient reconstruction of channel gain maps (CGMs) is essential for environment-aware wireless communications. Unlike prior works that focus on static CGMs at a single frequency band, this paper studies the construction of dynamic three-dimensional (3D) CGMs across multiple frequency bands over consecutive time slots, using limited channel measurements. Specifically, for a given transmitter (Tx) location, we represent the CGM at any given receiver (Rx) height as a two-dimensional (2D) grayscale image, where each pixel indicates the average channel gain over the corresponding Rx spatial grid cell. Accordingly, we propose ST-CGMDiff, a spatio-temporal latent diffusion framework that integrates sparse-observation conditioning and cross-frequency modeling to reconstruct dynamic 3D CGMs across multiple carrier frequencies and varying Rx heights within a unified model. In this framework, we jointly encode the input parameters, including the current sparse observations, observation masks, physical Tx and Rx \(z\)-coordinates, and carrier frequencies as spatial conditions. A frequency-conditioned convolutional gated recurrent unit (ConvGRU) then extracts causal information from previously constructed CGMs, while a spatially adaptive fusion mechanism balances historical information and current observations. Furthermore, bidirectional convolutional long short-term memory (BiConvLSTM) modules are introduced to capture the ordered dependencies among the selected Rx \(z\)-layers. Finally, we perform few-step latent diffusion followed by high-resolution residual refinement to generate the final CGM with improved spatial fidelity. Experiments on a dynamic multi-frequency dataset demonstrate that the proposed ST-CGMDiff consistently outperforms diffusion-based and conventional reconstruction baselines in reconstructing high-resolution CGMs from sparse observations. The proposed framework also achieves accurate contiguous-region completion with substantially lower inference cost, demonstrating its effectiveness for dynamic cross-frequency 3D CGM reconstruction.
\end{abstract}


\section{Introduction}

Sixth-generation (6G) wireless networks are envisioned to evolve beyond conventional connectivity toward the integration of communication, sensing, and intelligence, supporting emerging applications such as low-altitude aerial networks, embodied intelligent systems, and massive Internet-of-Things (IoT) deployments ~\cite{bHuawei6G,b6g2025,bembodied}. These applications rely on mobile terminals operating at different heights, dynamic propagation environments, and densely distributed devices, imposing increasingly stringent requirements on reliable connectivity and efficient network optimization. In particular, wireless channel conditions may vary significantly across locations, heights, time, and carrier frequencies due to blockage, reflections, and mobility. Therefore, conventional wireless communication design relying exclusively on instantaneous channel state information (CSI) requires repeated measurements and provides limited prior knowledge of channels at locations not yet visited by mobile terminals. These challenges motivate a paradigm shift toward environment-aware communications via channel knowledge maps (CKMs), where site-specific propagation knowledge is exploited to support proactive transmission and networking decisions~\cite{bZeng2021,b1}. Specifically, CKMs provide a structured representation of such knowledge by associating spatial locations with channel characteristics, thereby reducing the reliance on repeated channel acquisition and enabling environment-aware network optimization~\cite{b1,b2}.

Depending on the represented channel characteristics, CKMs can take different forms, such as CGMs, shadowing maps, beam index maps, and maps of multipath parameters~\cite{b1}. Among them, CGMs characterize the large-scale spatial variation of channel gains. Specifically, for a given Tx location, the considered service region can be discretized into spatial grid cells, each associated with the average channel gain over the corresponding Rx region. Such spatial averaging smooths fine-scale channel variations and reduces sensitivity to small positioning errors, making CGMs suitable for practical network deployment and optimization. In particular, spatially resolved channel-gain information can facilitate various wireless applications, including coverage optimization, user association, resource allocation, and mobility and trajectory planning~\cite{b1,b2,b3,b4}. The importance of spatial channel knowledge is further highlighted in unmanned aerial vehicle (UAV) communications, where CKM-based approaches have been investigated for positioning error compensation~\cite{b4} and multi-UAV placement optimization~\cite{bCKMMultiUAV2023}. However, conventional CGM designs mainly focus on 2D maps at a fixed Rx height, whereas aerial users may operate at varying heights in low-altitude wireless networks. Therefore, 3D CGMs are particularly important for characterizing channel variations in both horizontal and vertical dimensions~\cite{b3}. Nevertheless, constructing high-resolution 3D CGMs requires substantially more channel measurements and computational resources, especially in dynamic multi-frequency environments~\cite{b5}.

Existing CKM construction methods can be broadly categorized into model-based and measurement-driven approaches~\cite{b1,b2}. A representative model-based method is ray tracing, which reconstructs channel characteristics by simulating electromagnetic propagation based on environment geometry and material properties~\cite{bRT2024}. However, ray tracing requires detailed environmental descriptions and may incur high computational complexity for large-scale or high-resolution CKM construction ~\cite{b1,bRT2025}. Measurement-driven methods instead recover unobserved channel values directly from location-tagged measurements without requiring explicit environmental information. Classical approaches, such as kernel regression, Kriging, Gaussian-process regression, and matrix completion, exploit spatial correlation or low-rank structures ~\cite{b2,b5,b6,b7}. Other learning-based methods further capture nonlinear spatial relationships from sparse measurements~\cite{b8,b9,b10}. More recently, wireless radiance-field approaches based on neural radiance fields (NeRFs)~\cite{bNeRF} and 3D Gaussian splatting (3DGS)~\cite{b3DGS} have been explored for channel modeling and prediction. Representative works include NeRF$^2$~\cite{bNeRF2}, WRF-GS~\cite{bWRFGS}, and BiWGS~\cite{bBiWGS}, along with other recent advances~\cite{b11,b12,b13}. However, many radiance-field-based methods rely on scene-specific representation learning or adaptation, which may limit their generalization to unseen environments and introduce additional computational overhead ~\cite{bNeRF2,bGRaF}. For measurement-driven approaches, the data requirements for CGM construction and spatial channel prediction have also been analyzed in terms of sampling density and prediction accuracy~\cite{bCKMData2024}.

Recently, generative artificial intelligence (AI) has emerged as a promising paradigm for solving inverse problems with incomplete observations~\cite{b14}. Among various generative models, diffusion models have demonstrated strong capability in image synthesis, restoration, inpainting, and super-resolution, owing to their ability to learn complex data distributions and progressively recover structured signals from noise~\cite{b15}. These properties are particularly attractive for CKM reconstruction, where only sparse or partially missing channel measurements may be available and the underlying channel field exhibits complex spatial structures~\cite{b16,b17,b18}. For example, CKMDiff formulates CKM construction as a generative inverse problem and supports denoising, inpainting, and super-resolution without requiring explicit environment maps or base station (BS) locations~\cite{b8}. Subsequent studies extend diffusion-based CKM reconstruction to 3D scenarios. 3D-RadioDiff incorporates Rx height and environmental topology into the diffusion process~\cite{b19}, 3DCKMDiff jointly models multiple Rx \(z\)-layers for volumetric CKM reconstruction~\cite{b20}, and RadioDiff-3D employs 3D convolutional diffusion networks to jointly capture horizontal and vertical propagation structures ~\cite{b21}. Beyond diffusion models, linear transport guided flow matching (LT-GFM) has also been proposed for efficient CKM construction, covering both CGMs and spatial correlation maps~\cite{bGuidedFlow2026}.

Despite these advances, existing diffusion-based CKM reconstruction methods, including CKMDiff~\cite{b8}, 3D-RadioDiff~\cite{b19}, 3DCKMDiff~\cite{b20}, and RadioDiff-3D~\cite{b21}, primarily focus on static channel-map reconstruction without explicitly exploiting temporal correlations among consecutive channel states. Moreover, these methods are mainly developed for fixed carrier-frequency settings, without incorporating frequency-dependent propagation characteristics as conditioning information in a unified model. Regarding the vertical dimension, 3DCKMDiff~\cite{b20} jointly reconstructs multiple Rx $z$-layers, while RadioDiff-3D~\cite{b21} employs 3D convolutions for volumetric reconstruction. However, none of these diffusion models explicitly capture the ordered dependencies among flexibly selected Rx $z$-layers with varying physical $z$-coordinates. Furthermore, conventional iterative diffusion reconstruction may incur considerable inference costs~\cite{b22}.

To address these limitations, this paper proposes ST-CGMDiff, a sparse-conditioned latent diffusion framework for dynamic multi-frequency 3D CGM reconstruction. The framework combines frequency-aware spatial conditioning with spatio-temporal learning and efficient latent reconstruction. The main contributions are summarized as follows:
\begin{itemize}

\item We develop a unified cross-frequency 3D CGM reconstruction framework that explicitly incorporates carrier frequency and physical Tx and Rx \(z\)-coordinates as conditioning information. This enables a single trained ST-CGMDiff model to accommodate multiple carrier frequencies, varying Tx \(z\)-coordinates, and flexible Rx \(z\)-layer selections without requiring separate models for different frequencies or explicit environmental maps.

\item We propose a spatio-temporal reconstruction architecture that jointly exploits temporal and cross-\(z\)-layer dependencies. A frequency-conditioned ConvGRU extracts causal information from previous model reconstructions, while \(z\)-wise BiConvLSTM modules capture the ordered dependencies among Rx \(z\)-layers. A spatially adaptive fusion mechanism further balances the historical information with the current sparse observations, enabling closed-loop reconstruction over consecutive time slots.

\item We design an efficient few-step latent diffusion scheme with high-resolution residual refinement for both sparse-observation CGM super-resolution and contiguous-region CGM inpainting. Extensive experiments on a dynamic multi-frequency ray-tracing dataset show that the proposed ST-CGMDiff consistently outperforms the considered diffusion and conventional reconstruction baselines in sparse-observation super-resolution, achieves improved overall completion performance, and substantially reduces the inference cost compared with conventional diffusion-based reconstruction.

\end{itemize}

The remainder of this paper is organized as follows. Section II introduces the system model and formulates the dynamic multi-frequency 3D CGM reconstruction problem. Section III presents the proposed ST-CGMDiff framework, including spatial condition encoding, frequency-conditioned temporal fusion, height-aware latent diffusion, and residual refinement. Section IV describes the experimental setup and evaluates the reconstruction performance, cross-frequency generalization, and computational efficiency of ST-CGMDiff. Finally, Section V concludes the paper.

Notations: Boldface lower-case letters are used for vectors, while boldface upper-case letters are used for matrices and tensors. Calligraphic letters denote sets. $\mathbb{R}$ denotes the set of real numbers. The symbol $\odot$ denotes element-wise multiplication. $\|\cdot\|_1$ and $\|\cdot\|_2$ denote the $\ell_1$- and $\ell_2$-norms, respectively, and $\mathcal{N}(\mathbf{0},\mathbf{I})$ denotes the standard Gaussian distribution.

\section{System Model and Problem Formulation}

We consider a wireless network over a 3D spatial region. The propagation environment evolves over a finite time horizon consisting of \(T\) consecutive time slots due to the movement of scatterers and other objects, and the system operates over a set of carrier frequencies \(\mathcal{F}\). Our objective is to reconstruct the 3D CGMs over the considered region at different time slots and carrier frequency bands. Let \(\mathcal{T}=\{0,\ldots,T-1\}\) denote the set of time slots. At each time slot \(t\in\mathcal{T}\) and carrier frequency \(f\in\mathcal{F}\), the Tx and Rx are characterized by their 3D Cartesian coordinates.

\begin{figure}[!t]
\centering
\includegraphics[width=\columnwidth]{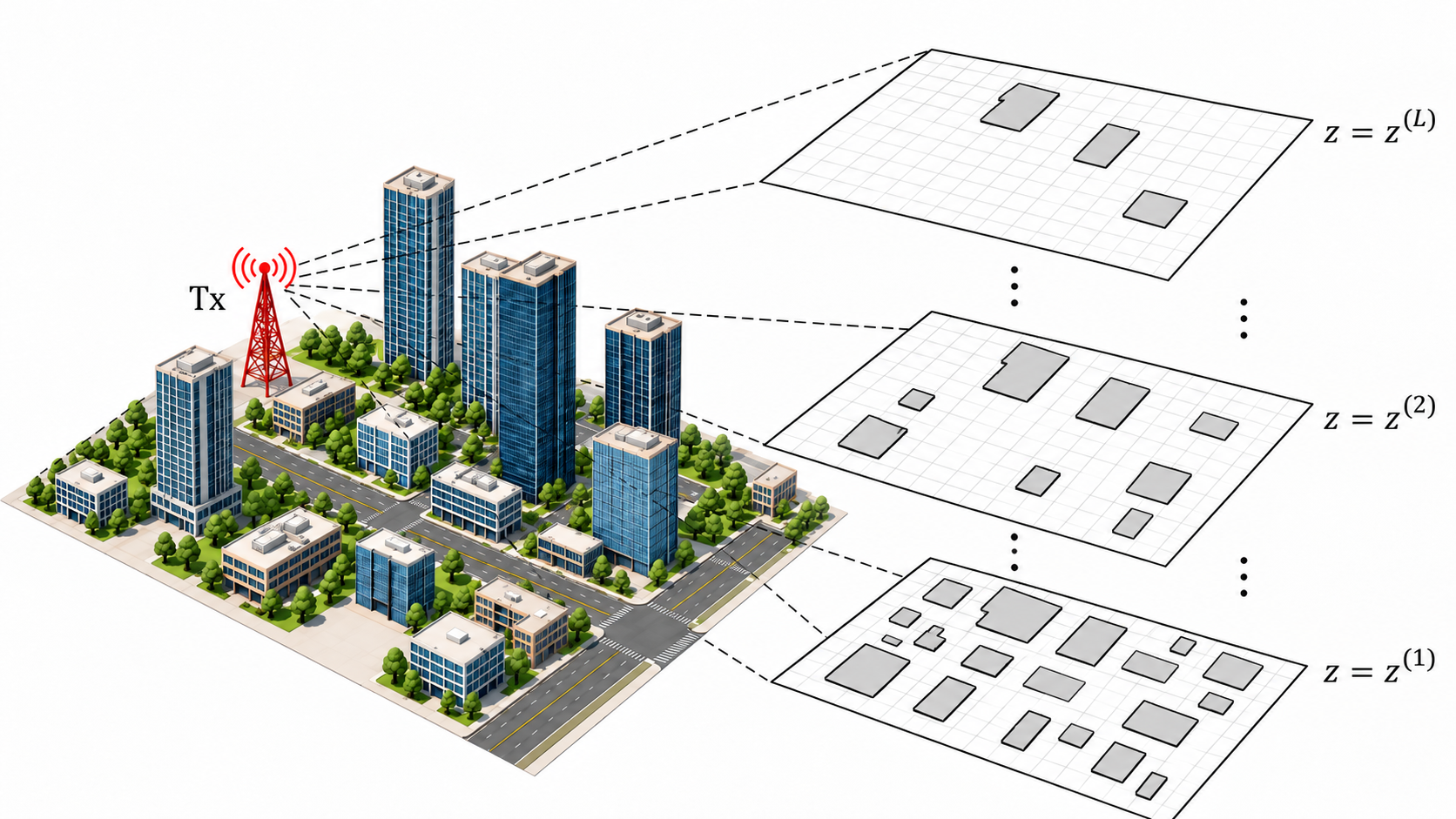}
\caption{Illustration of the considered 3D CGM reconstruction scenario. For a given Tx location, the Rx region is represented by horizontal planes at selected heights $z^{(1)},z^{(2)},\ldots,z^{(L)}$. Each plane corresponds to a 2D CGM to be reconstructed. The gray regions schematically indicate intersections of buildings with the respective horizontal planes, rather than channel-gain values.}
\label{fig:system_scenario}
\end{figure}

The reconstruction scenario is illustrated in Fig.~\ref{fig:system_scenario}.

Let
\begin{equation}
\mathbf{p}_{\mathrm{tx}} = (x_{\mathrm{tx}},y_{\mathrm{tx}},z_{\mathrm{tx}}) \in\mathbb{R}^{3}
\end{equation}
and
\begin{equation}
\mathbf{p}_{\mathrm{rx}} = (x_{\mathrm{rx}},y_{\mathrm{rx}},z_{\mathrm{rx}}) \in\mathbb{R}^{3}
\end{equation}
denote the Tx and Rx positions, respectively, where \((x,y)\) denotes the horizontal coordinates and \(z\) denotes the vertical coordinate.

At any given time slot \(t\) and carrier frequency \(f\), the channel gain between a Tx--Rx pair is described by
\begin{equation}
g_t: \mathbb{R}^{3} \times \mathbb{R}^{3} \times \mathcal{F} \rightarrow \mathbb{R},
\end{equation}
where \(g_t \left( \mathbf{p}_{\mathrm{tx}}, \mathbf{p}_{\mathrm{rx}};f \right)\) denotes the channel gain between the Tx at \(\mathbf{p}_{\mathrm{tx}}\) and the Rx at \(\mathbf{p}_{\mathrm{rx}}\). Therefore, the channel gain depends jointly on the 3D Tx and Rx positions $\mathbf{p}_{\mathrm{tx}}$ and $\mathbf{p}_{\mathrm{rx}}$, the time slot $t$, and the carrier frequency $f$.

To characterize the location-dependent channel information over the considered region, we adopt a grid-based CGM representation~\cite{bCGMScatterer2025}. At each Rx height, the horizontal service region is partitioned into an \(H\times W\) regular grid, where each grid cell corresponds to a small horizontal spatial region. The value associated with each grid cell is defined as the average channel gain over the corresponding region. Accordingly, each 2D CGM is represented as a single-channel grayscale image, where each pixel corresponds to one grid cell and its intensity represents the cell-averaged channel gain. Such a representation preserves the spatial organization and correlation of the channel field and facilitates reconstruction from sparse measurements and other environment-aware communication tasks ~\cite{b1,b2}.

For a given time slot \(t\), carrier frequency \(f\), and Rx height \(z^{(\ell)}\), we define the corresponding 2D CGM as
\begin{equation}
\mathbf{X}_{t,f}^{(\ell)} = \left[ X_{t,f}^{(\ell)}(m,n) \right]_{m=1,n=1}^{H,W} \in \mathbb{R}^{1\times H\times W},
\end{equation}
where the first dimension denotes the single channel of the grayscale CGM, and \(H\) and \(W\) denote the numbers of horizontal grid cells along the two spatial dimensions. The value of the \((m,n)\)-th grid cell is the average channel gain over the corresponding spatial region \(\mathcal{A}_{m,n}\)~\cite{bCGMScatterer2025}, i.e.,
\begin{equation}
X_{t,f}^{(\ell)}(m,n) = \frac{1}{|\mathcal{A}_{m,n}|} \int_{\mathcal{A}_{m,n}} g_t \left( \mathbf{p}_{\mathrm{tx}}, (x,y,z^{(\ell)});f \right) \,\mathrm{d}x\,\mathrm{d}y,
\end{equation}
where \(|\mathcal{A}_{m,n}|\) denotes the area of the \((m,n)\)-th grid cell.

Stacking the selected Rx \(z\)-layers gives the Rx-side 3D CGM
\begin{equation}
\mathbf{X}_{t,f} = \operatorname{Stack}_{\ell=1}^{L} \left( \mathbf{X}_{t,f}^{(\ell)} \right) \in \mathbb{R}^{L\times1\times H\times W},
\end{equation}
where \(\operatorname{Stack}_{\ell=1}^{L}(\cdot)\) denotes stacking the \(L\) 2D CGM layers along the Rx \(z\)-coordinate dimension. Thus, \(\mathbf{X}_{t,f}\) represents the grid-based 3D distribution of cell-averaged channel gains over the Rx region for a given Tx configuration, time slot \(t\), and carrier frequency \(f\).

\begin{figure*}[t]
\centering
\includegraphics[width=\textwidth]{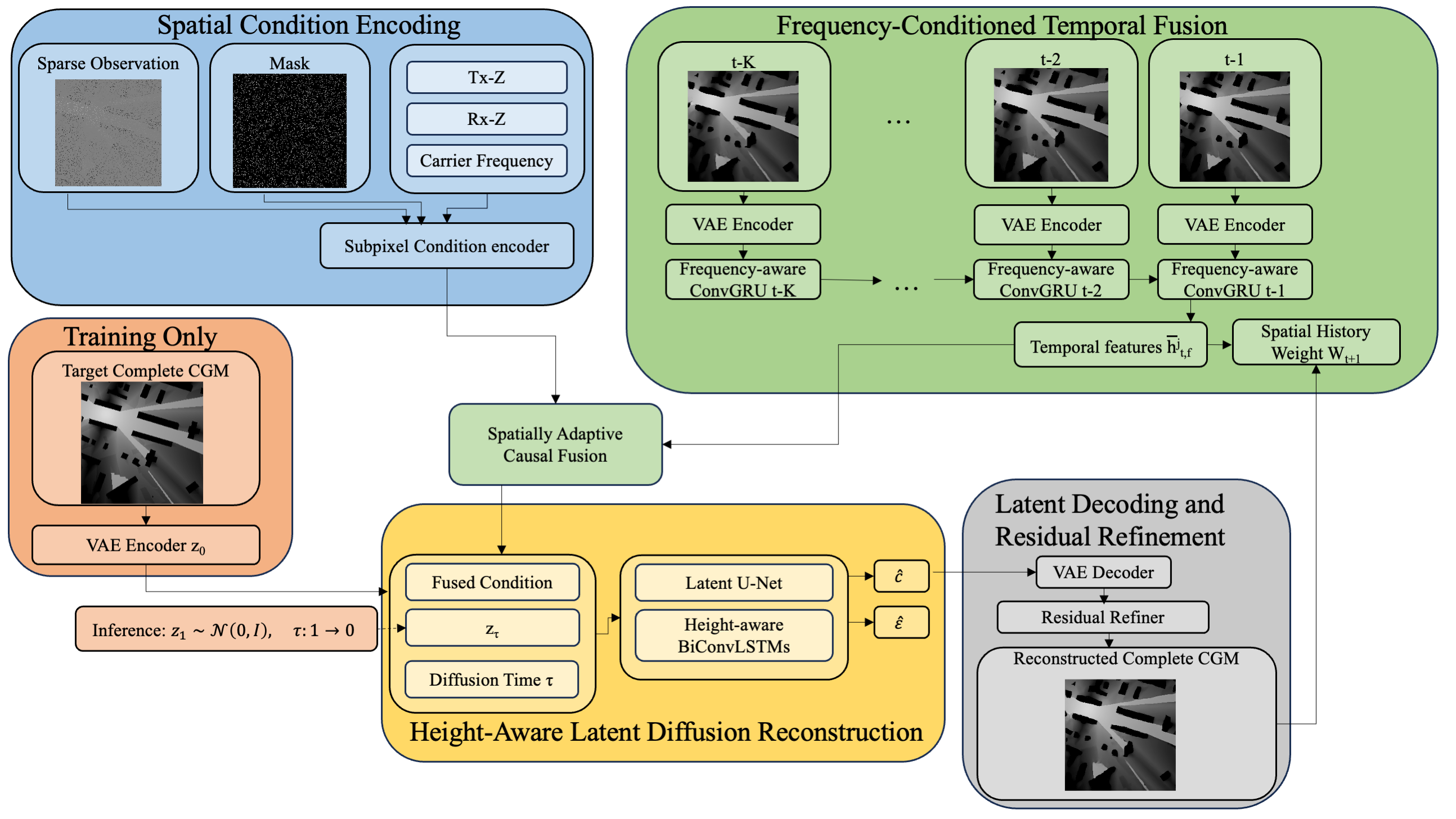}
\caption{Overall architecture of the proposed ST-CGMDiff. During training, the target CGM is encoded and perturbed at a randomly sampled diffusion time. During inference, reconstruction starts from Gaussian noise and is iteratively denoised from $\tau=1$ to $\tau=0$ under the fused spatial--temporal condition.}
\label{fig:overall}
\end{figure*}

Next, we introduce the channel-gain observations using a signed observation mask, defined as \(\mathbf{M}_{t,f} \in \{-1,1\}^{L\times1\times H\times W},\) where $1$ indicates an observed Rx grid cell and $-1$ denotes an unobserved grid cell. The corresponding sparse channel-gain observation is \(\mathbf{Y}_{t,f} = \frac{1+\mathbf{M}_{t,f}}{2} \odot \mathbf{X}_{t,f}.\)

The observed channel gains are assumed to be perfectly known, without measurement noise or channel-estimation errors. Therefore, whenever \([\mathbf{M}_{t,f}]_{\ell,1,m,n}=1\), the corresponding entry in \(\mathbf{Y}_{t,f}\) is exactly equal to the cell-averaged channel gain in \(\mathbf{X}_{t,f}\). Under this assumption, the reconstruction problem focuses on recovering the channel gains of the unobserved grid cells from the available spatial, frequency, and temporal information.

For a given time slot \(t\in\mathcal{T}\), the reconstruction targets are the complete 3D CGMs at all considered carrier frequencies, collected as
\begin{equation}
\mathcal{X}_{t} = \left\{ \mathbf{X}_{t,f} \right\}_{f\in\mathcal{F}}. \label{eq:current_cgm_collection}
\end{equation}
Here, \(\mathcal{X}_{t}\) is a frequency-indexed collection of unknown complete CGM volumes. The observation record available up to and including time slot \(t\) is
\begin{equation}
\mathcal{O}_{0:t} = \left\{ \left( \mathbf{Y}_{s,f}, \mathbf{M}_{s,f} \right) \right\}_{s=0,\ldots,t,\; f\in\mathcal{F}}. \label{eq:available_observations}
\end{equation}
\(\mathcal{O}_{0:t}\) contains the past and current observations, but no future observations. The reconstruction problem is to estimate \(\mathcal{X}_{t}\) from this record and the known physical conditions: the ordered list of Rx heights \(\mathcal{Z}=(z^{(1)},\ldots,z^{(L)})\), the Tx height \(z_{\mathrm{tx}}\), and the carrier frequencies in \(\mathcal{F}\).

In the proposed causal model, past observations are incorporated through an ordered history buffer of length \(K\) at the same carrier frequency \(f\):
\begin{equation}
\mathcal{P}_{t,f} = \left( \widehat{\mathbf{X}}_{t-K,f}, \ldots, \widehat{\mathbf{X}}_{t-1,f} \right). \label{eq:history_buffer}
\end{equation}
Here, \(K\) is a positive integer specifying the number of buffer entries, and \(\widehat{\mathbf{X}}_{s,f}\) denotes the final measurement-consistent reconstruction produced at time slot \(s\). The entries are ordered from oldest to newest. Only reconstructions already generated in the current closed-loop sequence are valid history entries; unavailable entries are represented by zero-tensor placeholders. A binary history-validity mask of length \(K\) assigns one to each available reconstruction and zero to each placeholder. At sequence startup, all \(K\) entries are invalid. This mask accompanies the buffer when it is supplied to the model. The model receives the current observation pair \((\mathbf{Y}_{t,f},\mathbf{M}_{t,f})\) directly and uses only the valid entries of \(\mathcal{P}_{t,f}\) for temporal modeling.

At any given time slot \(t\) and carrier frequency \(f\), channel-gain measurements are assumed to be available only for a subset of Rx grid cells. Each observation therefore represents the cell-averaged channel gain associated with a particular horizontal grid cell and Rx \(z\)-coordinate. These sparse observations are represented by \(\mathbf{Y}_{t,f}\) and \(\mathbf{M}_{t,f}\). Together with the selected Rx \(z\)-coordinates, Tx \(z\)-coordinate, carrier frequency, and causal historical CGMs, they constitute the information available for reconstruction.

Based on the current sparse channel-gain measurements and the available spatial, frequency, and temporal information, our objective is to reconstruct the complete grid-based 3D CGM over all selected Rx \(z\)-layers and all \(H\times W\) horizontal grid cells. For a given time slot \(t\in\mathcal{T}\) and carrier frequency \(f\in\mathcal{F}\), the unconstrained full-volume prediction is formulated as
\begin{equation}
\widetilde{\mathbf{X}}_{t,f} = F_{\theta} \left( \mathbf{Y}_{t,f}, \mathbf{M}_{t,f}, \mathcal{Z}, z_{\mathrm{tx}}, f, \mathcal{P}_{t,f} \right) \in \mathbb{R}^{L\times1\times H\times W},
\end{equation}
where \(F_{\theta}\) denotes the reconstruction model parameterized by \(\theta\).

The available channel-gain measurements are preserved in the final reconstruction according to
\begin{equation}
\widehat{\mathbf{X}}_{t,f} = \mathbf{Y}_{t,f} + \frac{1-\mathbf{M}_{t,f}}{2} \odot \widetilde{\mathbf{X}}_{t,f}, \label{eq:system_consistency}
\end{equation}
which guarantees the measurement-consistency condition
\begin{equation}
\frac{1+\mathbf{M}_{t,f}}{2} \odot \widehat{\mathbf{X}}_{t,f} = \mathbf{Y}_{t,f}.
\end{equation}

Depending on the spatial structure of the observation mask, the general reconstruction formulation contains two specific reconstruction tasks considered in this work as special cases, namely sparse-observation super-resolution and contiguous-region inpainting. In the sparse-observation super-resolution task, only a small and spatially distributed subset of CGM grid cells is observed, and the dense 3D CGM is reconstructed from these sparse cell-averaged channel-gain measurements. In the contiguous-region inpainting task, a connected group of horizontal grid cells is entirely unobserved, while the surrounding grid cells remain available, and the missing channel-gain region is reconstructed from the observed context.

\section{ST-CGMDiff for  Dynamic CGM Construction }

As illustrated in Fig.~\ref{fig:overall}, the proposed ST-CGMDiff reconstructs dynamic 3D CGMs through three main processing stages: Spatial Condition Encoding, Frequency-Conditioned Temporal Fusion, and Height-Aware Latent Diffusion and Decoding. First, the current sparse CGM observation, observation mask, Tx and Rx $z$-coordinates, and carrier frequency are jointly encoded to obtain the current frequency-aware spatial condition. Meanwhile, previously reconstructed CGMs are encoded into the latent space and processed by a frequency-conditioned ConvGRU to extract causal temporal information. The temporal representation is adaptively fused with the current spatial condition according to its spatial reliability. The resulting fused condition then guides a height-aware latent diffusion network, in which BiConvLSTM modules exchange information across the ordered Rx $z$-layers. Finally, the reconstructed latent representation is decoded into the spatial domain, refined at full resolution, and combined with the available observations through measurement consistency. The corresponding training objectives are introduced at the end of this section.

\subsection{Spatial Condition Encoding}

The signed observation mask \(\mathbf{M}_{t,f}\) defined in Section II is directly used for spatial condition encoding, with values \(1\) and \(-1\) indicating observed and unobserved grid cells, respectively.

The current spatial condition tensor is then constructed as
\begin{equation}
\mathbf{C}_{t,f} = \operatorname{Cat} \left[ \mathbf{Y}_{t,f}, \mathbf{M}_{t,f}, \mathbf{A}_{z,\mathrm{rx}}, \mathbf{A}_{z,\mathrm{tx}}, \mathbf{A}_{f} \right],
\end{equation}
with dimensions \(\mathbf{C}_{t,f} \in \mathbb{R}^{L\times5\times H\times W}.\) Here, \(\operatorname{Cat}[\cdot]\) denotes concatenation along the channel dimension. The tensors \(\mathbf{A}_{z,\mathrm{rx}}\), \(\mathbf{A}_{z,\mathrm{tx}}\), and \(\mathbf{A}_{f}\) contain the normalized Rx \(z\)-coordinates, Tx \(z\)-coordinate, and carrier frequency, respectively. They have dimensions \(L\times1\times H\times W\) after being broadcast over the corresponding Rx \(z\)-coordinate and \(x\)--\(y\) spatial dimensions. The Rx \(z\)-coordinate condition varies among the selected \(z\)-layers, whereas the Tx \(z\)-coordinate and carrier-frequency conditions are constant across the selected \(z\)-layers of the same CGM. By including these physical conditions alongside sparse observations, the encoder can form spatial features that reflect the selected Rx heights and carrier frequency within a shared model.

Let \(\mathbf{x}_{t,f}^{j}\), \(\mathbf{y}_{t,f}^{j}\), and \(\mathbf{m}_{t,f}^{j}\) denote the $j$th $z$-slices of \(\mathbf{X}_{t,f}\), \(\mathbf{Y}_{t,f}\), and \(\mathbf{M}_{t,f}\), respectively, where \(j\in\{1,\ldots,L\}\).

A subpixel condition encoder uses pixel unshuffle to reduce the horizontal spatial resolution of the sparse observation and signed observation mask while preserving their sampling locations. Let
\begin{equation}
\mathcal{P}_{r}: \mathbb{R}^{1\times H\times W} \rightarrow \mathbb{R}^{r^{2}\times H/r\times W/r}
\end{equation}
denote the pixel-unshuffle operation with spatial reduction factor \(r\). It rearranges the values in each \(r\times r\) spatial neighborhood into \(r^{2}\) feature channels without discarding information.

Let \(\mathbf{a}_{z,\mathrm{rx}}^{j}\), \(\mathbf{a}_{z,\mathrm{tx}}^{j}\), and \(\mathbf{a}_{f}^{j}\) denote the Rx $z$-coordinate, Tx $z$-coordinate, and frequency condition maps associated with the $j$th selected $z$-layer. These maps are broadcast at the spatial resolution of the encoded features. The spatial condition feature is
\begin{equation}
\mathbf{e}_{t,f}^{\mathrm{s},j} = E_{\mathrm{s}} \left( \operatorname{Cat} \left[ \mathcal{P}_{r} \left( \mathbf{y}_{t,f}^{j} \right), \mathcal{P}_{r} \left( \mathbf{m}_{t,f}^{j} \right), \mathbf{a}_{z,\mathrm{rx}}^{j}, \mathbf{a}_{z,\mathrm{tx}}^{j}, \mathbf{a}_{f}^{j} \right] \right), \label{eq:spatial_condition}
\end{equation}
where \(E_{\mathrm{s}}\) denotes the subpixel condition encoder and \(\mathbf{e}_{t,f}^{\mathrm{s},j}\) is the encoded current-condition feature for the $j$th Rx $z$-layer. The carrier frequency is normalized in the logarithmic domain before constructing \(\mathbf{a}_{f}^{j}\).

\subsection{Frequency-Conditioned Temporal Fusion}

For the \(k\)-th entry of the history buffer in \eqref{eq:history_buffer}, define its time index as \(\ell_k = t-K+k-1, \qquad k\in\{1,\ldots,K\}.\)

Let \(E_{\phi}\) denote the encoder of the pretrained variational autoencoder (VAE). The latent representation of the $j$th Rx $z$-layer in this history entry is
\begin{equation}
\mathbf{z}_{k,f}^{j} = \gamma E_{\phi} \left( \widehat{\mathbf{x}}_{\ell_k,f}^{j} \right), \qquad j\in\{1,\ldots,L\}, \label{eq:history_latent}
\end{equation}
where \(\widehat{\mathbf{x}}_{\ell_k,f}^{j}\) is the $j$th $z$-layer of \(\widehat{\mathbf{X}}_{\ell_k,f}\), \(\mathbf{z}_{k,f}^{j}\) is its latent representation, \(\phi\) denotes the pretrained VAE parameters, and \(\gamma\) is the latent scaling factor. The VAE remains frozen when the dynamic reconstruction model is trained.

For each Rx $z$-layer, the historical latent representations are processed chronologically by a stacked causal ConvGRU~\cite{b23}. Its gated updates aggregate information from the ordered history, while convolutional operations retain the horizontal spatial organization of each CGM layer. The recurrent hidden state is initialized to zero, and history entries marked as invalid by the history-validity mask do not contribute to the temporal representation. Suppressing the recurrent-layer index for clarity, one ConvGRU layer is expressed as
\begin{equation}
\mathbf{r}_{k,f}^{j} = \sigma \left( \mathbf{W}_{\mathrm r} * \operatorname{Cat} \left[ \mathbf{z}_{k,f}^{j}, \mathbf{h}_{k-1,f}^{j} \right] + \mathbf{b}_{\mathrm r} \right), \label{eq:convgru_reset}
\end{equation}

\begin{equation}
\mathbf{u}_{k,f}^{j} = \sigma \left( \mathbf{W}_{\mathrm u} * \operatorname{Cat} \left[ \mathbf{z}_{k,f}^{j}, \mathbf{h}_{k-1,f}^{j} \right] + \mathbf{b}_{\mathrm u} \right), \label{eq:convgru_update}
\end{equation}

\begin{equation}
\widetilde{\mathbf{h}}_{k,f}^{j} = \tanh \left( \mathbf{W}_{\mathrm h} * \operatorname{Cat} \left[ \mathbf{z}_{k,f}^{j}, \mathbf{r}_{k,f}^{j} \odot \mathbf{h}_{k-1,f}^{j} \right] + \mathbf{b}_{\mathrm h} \right), \label{eq:convgru_candidate}
\end{equation}

\begin{equation}
\mathbf{h}_{k,f}^{j} = \mathbf{u}_{k,f}^{j} \odot \mathbf{h}_{k-1,f}^{j} + \left( 1-\mathbf{u}_{k,f}^{j} \right) \odot \widetilde{\mathbf{h}}_{k,f}^{j}. \label{eq:convgru_hidden}
\end{equation}

Here, \(\mathbf{h}_{0,f}^{j}=\mathbf{0}\); \(\mathbf{r}_{k,f}^{j}\) and \(\mathbf{u}_{k,f}^{j}\) are the reset and update gates; \(\widetilde{\mathbf{h}}_{k,f}^{j}\) is the candidate hidden state; and \(\mathbf{h}_{k,f}^{j}\) is the updated hidden state. The tensors \(\mathbf{W}_{\mathrm r}\), \(\mathbf{W}_{\mathrm u}\), and \(\mathbf{W}_{\mathrm h}\) are convolution kernels, while \(\mathbf{b}_{\mathrm r}\), \(\mathbf{b}_{\mathrm u}\), and \(\mathbf{b}_{\mathrm h}\) are the corresponding biases. Moreover, \(*\) denotes convolution, \(\sigma(\cdot)\) is the sigmoid function, and \(\tanh(\cdot)\) is the hyperbolic tangent function. In the stacked implementation, the hidden state produced by one ConvGRU layer is provided to the next layer. The final state \(\mathbf{h}_{K,f}^{j}\) summarizes the historical evolution of the $j$th Rx $z$-layer.

The ConvGRU representation is further conditioned on carrier frequency. This permits the temporal features to vary with the frequency-dependent propagation characteristics without using separate temporal networks for different carrier frequencies. A frequency-conditioning network produces \(\left( \boldsymbol{\alpha}_{f}, \boldsymbol{\beta}_{f} \right) = H_f \left( \overline{f} \right),\) and applies feature-wise modulation as
\begin{equation}
\overline{\mathbf{h}}_{t,f}^{j} = \left( 1+\boldsymbol{\alpha}_{f} \right) \odot \mathbf{h}_{K,f}^{j} + \boldsymbol{\beta}_{f}. \label{eq:frequency_film}
\end{equation}

Here, \(\overline f\) is the normalized scalar carrier frequency used to construct \(\mathbf{A}_{f}\), \(H_f(\cdot)\) is a learnable frequency-conditioning network, and \(\boldsymbol{\alpha}_{f}\) and \(\boldsymbol{\beta}_{f}\) are channel-wise scaling and shifting parameters, respectively. The same modulation parameters are applied to all selected Rx $z$-layers and broadcast over their spatial dimensions. The resulting \(\overline{\mathbf{h}}_{t,f}^{j}\) is the frequency-conditioned temporal representation of the $j$th $z$-layer.

Because the history buffer contains model reconstructions rather than ground truth, historical features need not be equally informative at all spatial locations. We therefore adaptively weight temporal features instead of treating their contribution as spatially uniform. The temporal representation is projected into the feature space of the current spatial condition and fused as
\begin{equation}
\mathbf{e}_{t,f}^{\mathrm{fuse},j} = \mathbf{e}_{t,f}^{\mathrm{s},j} + \mathbf{W}_{t,f}^{j} \odot \sigma \left[ G_{\mathrm t} \left( \overline{\mathbf{h}}_{t,f}^{j} \right) \right] \odot P_{\mathrm t} \left( \overline{\mathbf{h}}_{t,f}^{j} \right). \label{eq:causal_fusion}
\end{equation}

Here, \(P_{\mathrm t}(\cdot)\) projects the temporal representation into the spatial-condition feature space, \(G_{\mathrm t}(\cdot)\) generates a learnable fusion gate, and \(\mathbf{W}_{t,f}^{j}\) is the spatial history weight calculated after reconstructing the preceding frame and resized to the feature resolution. The current spatial condition \(\mathbf{e}_{t,f}^{\mathrm{s},j}\) retains unit weight, while the fusion gate and spatial history weight jointly control the contribution of the temporal representation. The resulting \(\mathbf{e}_{t,f}^{\mathrm{fuse},j}\) is supplied to the latent diffusion network.

The frequency-conditioned temporal features also produce a current-frame prediction:
\begin{equation}
\widehat{\mathbf{X}}_{t,f}^{\mathrm{temp}} = T_{\eta} \left( \left\{ \overline{\mathbf{h}}_{t,f}^{j} \right\}_{j=1}^{L} \right), \label{eq:temporal_prediction}
\end{equation}
where \(T_{\eta}\) denotes the temporal prediction head parameterized by \(\eta\), and \(\widehat{\mathbf{X}}_{t,f}^{\mathrm{temp}}\) is the complete CGM predicted solely from the historical temporal representation.

After the final reconstruction \(\widehat{\mathbf{X}}_{t,f}\) is obtained, the local structural similarity index measure (SSIM) map and next-frame history weight are calculated as
\begin{align}
\mathbf{S}_{t,f} &= \operatorname{SSIM}_{\mathrm{loc}} \left( \widehat{\mathbf{X}}_{t,f}^{\mathrm{temp}}, \widehat{\mathbf{X}}_{t,f} \right),\\
\mathbf{W}_{t+1,f} &= \Psi \left( \mathbf{S}_{t,f} \right), \label{eq:history_feedback}
\end{align}
where \(\operatorname{SSIM}_{\mathrm{loc}}(\cdot,\cdot)\) denotes the local SSIM operation, \(\mathbf{S}_{t,f}\) is the resulting spatial similarity map, and \(\Psi(\cdot)\) denotes the normalization and clipping operation that converts the similarity values into a valid history-weight map. The $j$th $z$-layer slice of this map is denoted by \(\mathbf{W}_{t+1,f}^{j}\). A high local similarity assigns greater weight to the historical context at the next time slot, whereas a low similarity makes the model rely primarily on the new sparse observation.

At sequence startup, the history buffer is filled with zero tensors, the corresponding entries are marked as invalid, and the spatial history weight is initialized to zero. The first frame is therefore reconstructed from the current spatial condition. During closed-loop training and inference, each model reconstruction is appended to the history buffer and the oldest entry is discarded. Ground-truth historical CGMs are not supplied to the temporal module.

\subsection{Height-Aware Latent Diffusion and Decoding}

The fused spatial-temporal condition is used to guide the reconstruction of the current CGM in a compact latent space~\cite{b22,b24}. Performing diffusion in this representation reduces the spatial resolution of the iterative denoising process compared with operating directly on the full-resolution CGM. During training, each target CGM slice is first encoded by the frozen VAE. Stacking the latent representations of the selected Rx $z$-coordinates gives
\begin{equation}
\mathbf{z}_{0,t,f} = \operatorname{Stack}_{j=1}^{L} \left[ \gamma E_{\phi} \left( \mathbf{x}_{t,f}^{j} \right) \right]. \label{eq:target_latent}
\end{equation}
Here, \(\mathbf{z}_{0,t,f}\) is the clean latent representation of the current complete CGM. To distinguish diffusion time from the physical time index \(t\), the diffusion variable is denoted by \(\tau\). The forward diffusion process~\cite{b24} is
\begin{equation}
\mathbf{z}_{\tau,t,f} = (1-\tau)\mathbf{z}_{0,t,f} + \sqrt{\tau}\boldsymbol{\epsilon}, \qquad \boldsymbol{\epsilon} \sim \mathcal{N} \left( \mathbf{0}, \mathbf{I} \right), \label{eq:forward_diffusion}
\end{equation}
where \(\tau\) is randomly sampled during training, \(\boldsymbol{\epsilon}\) is standard Gaussian noise, and \(\mathbf{0}\) and \(\mathbf{I}\) denote the zero mean vector and identity covariance matrix, respectively.

For the $j$th Rx $z$-layer, the noisy latent feature is concatenated with its fused condition:
\begin{equation}
\mathbf{q}_{\tau,t,f}^{j} = \operatorname{Cat} \left[ \mathbf{z}_{\tau,t,f}^{j}, \mathbf{e}_{t,f}^{\mathrm{fuse},j} \right], \label{eq:unet_input}
\end{equation}
where \(\mathbf{z}_{\tau,t,f}^{j}\) is the $j$th $z$-slice of \(\mathbf{z}_{\tau,t,f}\), and \(\mathbf{q}_{\tau,t,f}^{j}\) is the corresponding input feature of the latent U-Net. The U-Net processes horizontal spatial features slice by slice together with an embedding of \(\tau\), while embedded BiConvLSTM modules exchange information across the ordered Rx $z$-coordinate dimension.

\begin{figure}[t]
\centering
\includegraphics[width=\columnwidth]{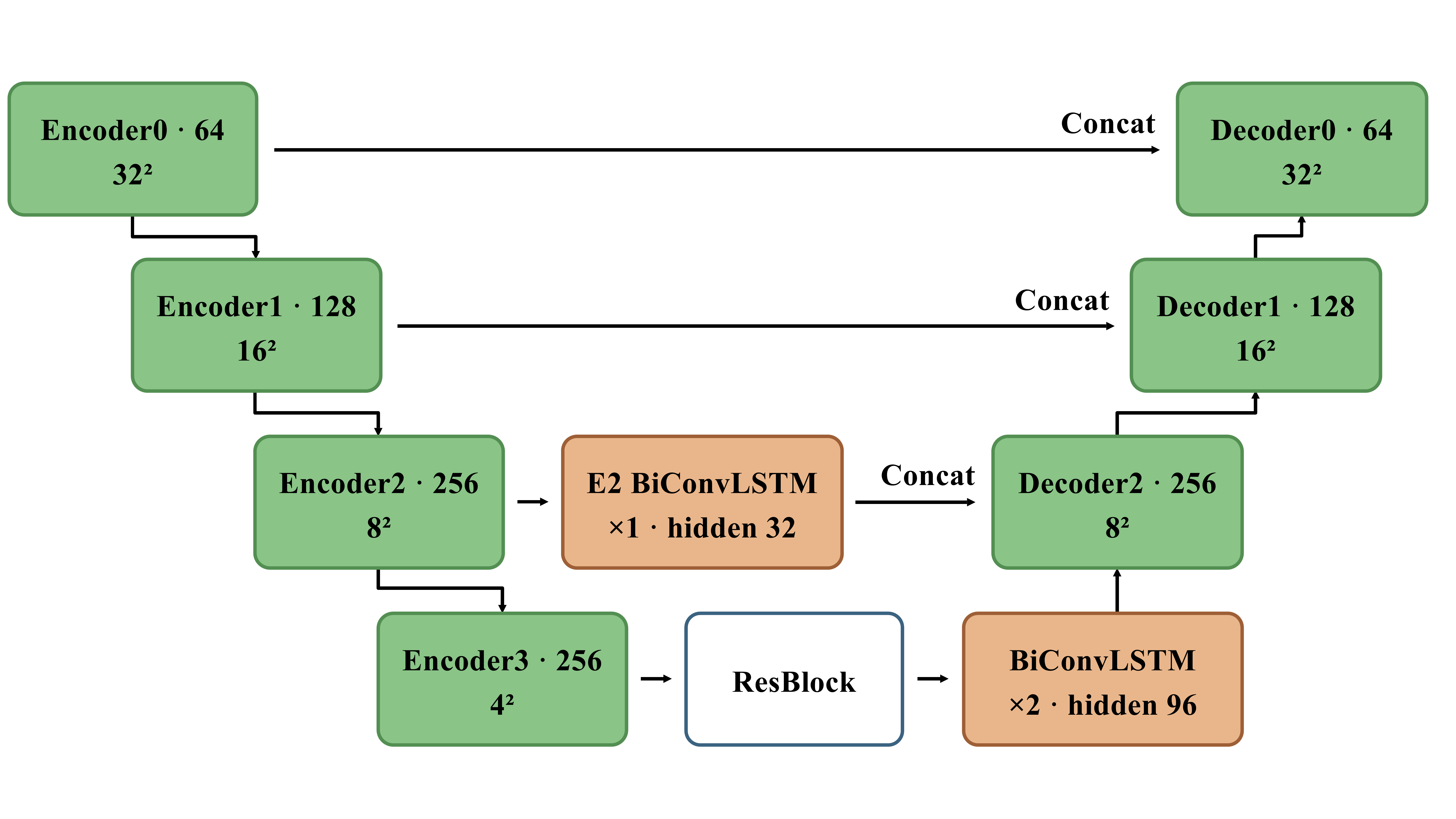}
\caption{Height-aware latent reconstruction network. The U-Net extracts horizontal spatial features, while the embedded BiConvLSTM modules exchange information among the ordered Rx $z$-layers.}
\label{fig:unet}
\end{figure}

The forward and backward recurrent directions allow each selected $z$-layer to receive information from both lower and higher Rx $z$-layers. Unlike independently processing each Rx \(z\)-layer, BiConvLSTM exchanges cross-layer information through convolutional recurrent operations while preserving the horizontal spatial layout. Because the physical Rx $z$-coordinates are explicitly included in the spatial condition, the model can process different numbers and selections of Rx $z$-layers without assuming uniformly spaced $z$-coordinates.

The $z$-aware latent U-Net \(U_{\theta}\) produces an attenuation prediction and a noise prediction:
\begin{equation}
\left( \mathbf{c}_{\theta}, \boldsymbol{\epsilon}_{\theta} \right) = U_{\theta} \left( \mathbf{z}_{\tau,t,f}, \mathbf{e}_{t,f}^{\mathrm{fuse}}, \tau \right), \label{eq:unet_outputs}
\end{equation}
where
\begin{equation}
\mathbf{e}_{t,f}^{\mathrm{fuse}} = \operatorname{Stack}_{j=1}^{L} \left( \mathbf{e}_{t,f}^{\mathrm{fuse},j} \right).
\end{equation}

Here, \(\mathbf{c}_{\theta}\) and \(\boldsymbol{\epsilon}_{\theta}\) are the predicted attenuation and noise terms, respectively. Since the attenuation head is supervised toward \(-\mathbf{z}_{0,t,f}\), the clean latent estimate is
\begin{equation}
\widehat{\mathbf{z}}_{0,t,f} = -\mathbf{c}_{\theta}. \label{eq:clean_latent}
\end{equation}

During inference, the reverse diffusion process starts from Gaussian noise and uses a reduced sequence of denoising steps to obtain \(\widehat{\mathbf{z}}_{0,t,f}\). The reconstructed latent is then mapped back to the spatial domain using the frozen VAE decoder \(D_{\phi}\). The coarse CGM is obtained as
\begin{equation}
\mathbf{X}_{t,f}^{\mathrm{c}} = D_{\phi} \left( \frac{ \widehat{\mathbf{z}}_{0,t,f} }{ \gamma } \right). \label{eq:coarse_reconstruction}
\end{equation}

Since latent-space reconstruction may lose fine-scale spatial details, a full-resolution residual refiner is employed to correct the coarse CGM using both the decoded reconstruction and the current spatial conditions:
\begin{equation}
\Delta\mathbf{X}_{t,f} = R_{\omega} \left( \operatorname{Cat} \left[ \mathbf{X}_{t,f}^{\mathrm{c}}, \mathbf{Y}_{t,f}, \mathbf{M}_{t,f}, \mathbf{A}_{z,\mathrm{rx}}, \mathbf{A}_{z,\mathrm{tx}}, \mathbf{A}_{f} \right] \right), \label{eq:residual_correction}
\end{equation}
and the refined CGM is obtained as
\begin{equation}
\widetilde{\mathbf{X}}_{t,f} = \mathbf{X}_{t,f}^{\mathrm{c}} + \Delta\mathbf{X}_{t,f}. \label{eq:refined_prediction}
\end{equation}

Here, \(R_{\omega}\) denotes the residual refiner parameterized by \(\omega\), \(\Delta\mathbf{X}_{t,f}\) is the predicted residual correction, and \(\widetilde{\mathbf{X}}_{t,f}\) is the refined CGM before observation consistency is imposed. The refiner operates at the original spatial resolution and exchanges information across adjacent Rx $z$-layers.

Finally, observation consistency preserves the available channel-gain measurements:
\begin{equation}
\widehat{\mathbf{X}}_{t,f} = \mathbf{Y}_{t,f} + \frac{1-\mathbf{M}_{t,f}}{2} \odot \widetilde{\mathbf{X}}_{t,f}. \label{eq:observation_consistency}
\end{equation}
Therefore, the measured channel gains are retained at observed positions, while the model supplies reconstructed values only at unobserved positions.

Compared with existing diffusion-based CKM reconstruction methods, including CKMDiff~\cite{b8}, 3D-RadioDiff~\cite{b19}, 3DCKMDiff~\cite{b20}, and RadioDiff-3D~\cite{b21}, the proposed ST-CGMDiff differs in several key architectural aspects. First, rather than reconstructing individual static CGMs, ST-CGMDiff incorporates a frequency-conditioned ConvGRU and spatially adaptive fusion mechanism to exploit causal information from previous reconstructions. Second, carrier frequency and physical Tx and Rx $z$-coordinates are explicitly incorporated as conditioning information, enabling a unified model to handle multiple frequency bands and flexible Rx-height selections. Third, unlike existing 3D approaches that mainly rely on Rx-height conditioning or 3D convolutions, the proposed architecture employs $z$-wise BiConvLSTM modules to explicitly model the ordered dependencies among Rx $z$-layers. Finally, reconstruction is performed through few-step diffusion in a compact latent space, followed by full-resolution residual refinement, which reduces the iterative reconstruction burden while recovering fine spatial details.

\subsection{Loss Function}
To train the proposed ST-CGMDiff framework, we jointly optimize the latent-space diffusion objective, pixel-domain reconstruction objective, and temporal-prediction objective. We first define the latent-space training objective, which supervises the attenuation prediction, noise prediction, and clean latent reconstruction:
\begin{align}
\mathcal{L}_{\mathrm{att}} &= \left\| \mathbf{c}_{\theta} + \mathbf{z}_{0,t,f} \right\|_{2}^{2},\\
\mathcal{L}_{\mathrm{noise}} &= \left\| \boldsymbol{\epsilon}_{\theta} - \boldsymbol{\epsilon} \right\|_{2}^{2},\\
\mathcal{L}_{\mathrm{clean}} &= \left\| \widehat{\mathbf{z}}_{0,t,f} - \mathbf{z}_{0,t,f} \right\|_{1}.
\end{align}

The complete latent objective is
\begin{equation}
\mathcal{L}_{\mathrm{lat}} = \mathcal{L}_{\mathrm{att}} + \mathcal{L}_{\mathrm{noise}} + \lambda_{\mathrm{clean}} \mathcal{L}_{\mathrm{clean}}, \label{eq:latent_loss}
\end{equation}
where \(\lambda_{\mathrm{clean}}\) controls the contribution of the clean-latent reconstruction loss.

Let \(\boldsymbol{\Omega}_{t,f} = \frac{1-\mathbf{M}_{t,f}}{2}\) denote the missing-region mask. The reconstruction loss over the unobserved positions is
\begin{equation}
\mathcal{L}_{\mathrm{miss}} = \frac{ \left\| \boldsymbol{\Omega}_{t,f} \odot \left( \widehat{\mathbf{X}}_{t,f} - \mathbf{X}_{t,f} \right) \right\|_{1} }{ \left\| \boldsymbol{\Omega}_{t,f} \right\|_{1} + \delta }, \label{eq:missing_loss}
\end{equation}
where \(\delta\) is a small positive constant used for numerical stability.

A spatial-gradient loss preserves local transitions:
\begin{equation}
\mathcal{L}_{\mathrm{grad}} = \frac{ \displaystyle \sum_{d\in\{x,y\}} \left\| \boldsymbol{\Omega}_{t,f} \odot \left( \nabla_d\widehat{\mathbf{X}}_{t,f} - \nabla_d\mathbf{X}_{t,f} \right) \right\|_{1} }{ \left\| \boldsymbol{\Omega}_{t,f} \right\|_{1} + \delta }, \label{eq:gradient_loss}
\end{equation}
where \(\nabla_d\) denotes the spatial finite-difference operator along the horizontal direction \(d\in\{x,y\}\).

A multiscale reconstruction loss is defined as
\begin{equation}
\mathcal{L}_{\mathrm{ms}} = \sum_{q\in\mathcal{Q}} \frac{ \left\| D_q \left( \boldsymbol{\Omega}_{t,f} \odot \widehat{\mathbf{X}}_{t,f} \right) - D_q \left( \boldsymbol{\Omega}_{t,f} \odot \mathbf{X}_{t,f} \right) \right\|_{1} }{ \left\| D_q \left( \boldsymbol{\Omega}_{t,f} \right) \right\|_{1} + \delta }, \label{eq:multiscale_loss}
\end{equation}
where \(\mathcal{Q}\) is the set of considered spatial scales and \(D_q(\cdot)\) denotes the downsampling operator associated with \(q\in\mathcal{Q}\).

Structural similarity is encouraged by
\begin{equation}
\mathcal{L}_{\mathrm{ssim}} = 1- \operatorname{SSIM} \left( \widehat{\mathbf{X}}_{t,f}, \mathbf{X}_{t,f} \right), \label{eq:ssim_loss}
\end{equation}
where \(\operatorname{SSIM}(\cdot,\cdot)\) denotes the structural similarity between the reconstructed and target CGM volumes.

The pixel-domain objective is
\begin{equation}
\mathcal{L}_{\mathrm{pix}} = \mathcal{L}_{\mathrm{miss}} + \lambda_{\mathrm{grad}} \mathcal{L}_{\mathrm{grad}} + \lambda_{\mathrm{ms}} \mathcal{L}_{\mathrm{ms}} + \lambda_{\mathrm{ssim}} \mathcal{L}_{\mathrm{ssim}}, \label{eq:pixel_loss}
\end{equation}
where \(\lambda_{\mathrm{grad}}\), \(\lambda_{\mathrm{ms}}\), and \(\lambda_{\mathrm{ssim}}\) weight the spatial-gradient, multiscale, and structural-similarity losses, respectively.

To ensure that the history-only prediction provides a meaningful reference for causal feedback, the temporal prediction head is supervised independently:
\begin{equation}
\mathcal{L}_{\mathrm{pred}} = \frac{ \left\| \boldsymbol{\Omega}_{t,f} \odot \left( \widehat{\mathbf{X}}_{t,f}^{\mathrm{temp}} - \mathbf{X}_{t,f} \right) \right\|_{1} }{ \left\| \boldsymbol{\Omega}_{t,f} \right\|_{1} + \delta }. \label{eq:temporal_prediction_loss}
\end{equation}

The complete training objective is
\begin{equation}
\mathcal{L} = \mathcal{L}_{\mathrm{lat}} + \lambda_{\mathrm{pix}} \mathcal{L}_{\mathrm{pix}} + \lambda_{\mathrm{pred}} \mathcal{L}_{\mathrm{pred}}, \label{eq:total_loss}
\end{equation}
where \(\lambda_{\mathrm{pix}}\) and \(\lambda_{\mathrm{pred}}\) control the pixel-domain reconstruction and temporal-prediction objectives, respectively. The values of all loss weights are provided in the experimental settings.

\section{Results and Analysis}
\label{sec:results}

We evaluate sparse-observation CGM super-resolution and contiguous-region completion. Both dynamic super-resolution and completion use zero-start closed-loop inference with previous reconstructions as history. Root mean square error (RMSE), normalized mean square error (NMSE), and peak signal-to-noise ratio (PSNR) are evaluated at unobserved positions, with SSIM evaluation regions specified below.

\subsection{Dataset Preparation}
\label{subsec:dataset}

The dataset uses Sionna RT~\cite{b25} and all 701 UrbanRadio3D building layouts~\cite{b21}. Four Tx configurations per layout share their geometry and dynamic trajectories across four carrier frequencies; Table~\ref{tab:dataset_config} lists the ray-tracing settings. The buildings use the International Telecommunication Union (ITU) concrete material model.

\begin{table}[t]
\centering
\caption{Dataset and Ray-Tracing Configuration.}
\label{tab:dataset_config}
\footnotesize
\renewcommand{\arraystretch}{1.08}
\setlength{\tabcolsep}{3.2pt}
\begin{tabular}{@{}p{0.49\columnwidth}p{0.41\columnwidth}@{}}
\toprule
\textbf{Parameter} & \textbf{Setting} \\
\midrule
Building layouts & 701 \\
Scene size & $256\,\mathrm{m}\times256\,\mathrm{m}$ \\
Horizontal grid & $256\times256$ \\
Spatial resolution & $1\,\mathrm{m}$ \\
Tx configurations per layout & 4 \\
Training frequencies &
$2.4,\,3.5,\,5.9,\,28\,\mathrm{GHz}$ \\
Unseen test frequencies &
$4.7,\,15.1\,\mathrm{GHz}$ \\
Tx height & $1.5$--$40\,\mathrm{m}$ \\
Rx height range & $1$--$40\,\mathrm{m}$ \\
Number of Rx \(z\)-layers & 50 \\
Frames per sequence & 15 \\
Ground-level scatterers & 18 \\
Aerial scatterers & 3 \\
Aerial-scatterer height & $15$--$40\,\mathrm{m}$ \\
Propagation mechanisms &
Direct, reflection, diffraction \\
Diffuse scattering & Disabled \\
Maximum propagation depth & 1 \\
Rays per Tx--frequency pair &
$2\times10^{8}$ \\
Building material & ITU concrete \\
Building relative permittivity & 5.24 \\
Ground relative permittivity & 15 \\
Ground conductivity & $0.005\,\mathrm{S/m}$ \\
Shadow-blending coefficient & 0.25 \\
Train/validation/test layouts & 600 / 50 / 51 \\
Channel-gain range &
$[-190,-70]\,\mathrm{dB}$ \\
\bottomrule
\end{tabular}
\end{table}

Each Tx--frequency pair yields a dynamic sequence: 16 per layout and 11,216 in total, each with 15 frames and 50 Rx \(z\)-layers. The Rx heights span $1$--$40\,\mathrm{m}$, comprising the endpoints and 48 sorted stratified-random interior heights. Physical Tx/Rx heights and carrier frequency are explicit conditions, not inferred from spatial or layer indices.

Scenes contain ground-level and aerial scatterers. To limit ray-tracing cost while retaining dominant deterministic components, diffuse scattering is disabled and propagation depth is one, with $N_{\mathrm{ray}}=2\times10^{8}$ rays per Tx--frequency pair.

For ITU concrete~\cite{b26}, the frequency-dependent conductivity is \(\sigma(f)=0.0462 f^{0.7822}\ \mathrm{S/m},\) where $f$ is in GHz. At $2.4$, $3.5$, $5.9$, and $28\,\mathrm{GHz}$, it equals $0.09163$, $0.12309$, $0.18518$, and $0.62605\,\mathrm{S/m}$, respectively.

The dataset is divided by building layout such that no layout is shared among the training, validation, and test subsets, which contain 600, 50, and 51 layouts, respectively. The complete four-frequency dataset contains 168,240 3D CGM frames and 8.412 million horizontal CGM slices, with each training frequency contributing 2,804 sequences.

To evaluate cross-frequency generalization, two additional test sets are generated at 4.7 and 15.1~GHz using the same 51 test layouts, Tx configurations, scene geometries, dynamic trajectories, and ray-tracing settings. These frequencies are not used for training, validation, checkpoint selection, or model adaptation. Each unseen frequency contains 204 sequences, corresponding to 3,060 3D CGM frames and 153,000 horizontal CGM slices. Their ITU-concrete conductivities are 0.15501 and $0.38622\,\mathrm{S/m}$, respectively.

Channel gains are clipped to $[-190,-70]\,\mathrm{dB}$ and linearly normalized to $[-1,1]$ for training and inference.

\subsection{Experimental Setup}
\label{subsec:experimental_setup}

Table~\ref{tab:model_config} summarizes task sampling and model configurations. The feedback settings include an exponential moving average (EMA) coefficient. During training and evaluation, 10--50 Rx $z$-layers are randomly selected, ordered by height, and supplied with their physical coordinates. Super-resolution samples each layer uniformly at an independent $5\%$--$10\%$ rate. Completion masks an $81\times81$ square ($\approx 10.01\%$ of the $256\times256$ grid) per layer; its location is shared across selected layers and frames of a sequence, but resampled for each layout and Tx configuration. Only the unknown region is evaluated.

\begin{table}[t]
\centering
\caption{Task and Model Configuration.}
\label{tab:model_config}
\footnotesize
\renewcommand{\arraystretch}{1.08}
\setlength{\tabcolsep}{3.2pt}
\begin{tabular}{@{}p{0.51\columnwidth}p{0.39\columnwidth}@{}}
\toprule
\textbf{Parameter} & \textbf{Setting} \\
\midrule
Selected Rx \(z\)-layers & 10--50 \\
Super-resolution sampling rate & $5\%$--$10\%$ \\
Completion region & $81\times81$ \\
Condition channels & 5 \\
Observation-mask values &
$\{-1,1\}$ \\
Tx/Rx height normalization & Linear \\
Frequency normalization & Logarithmic \\
VAE latent channels & 4 \\
VAE base channels & 64 \\
Latent scaling factor \(\gamma\) & 0.362 \\
Condition reduction factor \(r\) & 8 \\
Condition feature resolution & $32\times32$ \\
Condition feature channels & 32 \\
Latent U-Net base channels & 64 \\
Latent U-Net hidden channels & 96 \\
BiConvLSTM layers & 2 \\
ConvGRU layers & 4 \\
ConvGRU hidden channels & 64 \\
ConvGRU kernel size & $3\times3$ \\
History length \(K\) & 4 \\
Feedback sigmoid slope / center & 12 / 0.8 \\
Feedback EMA coefficient & 0.7 \\
Reverse diffusion steps & 10 \\
\bottomrule
\end{tabular}
\end{table}

The five spatial-conditioning inputs are the observed CGM, signed observation mask, Rx and Tx heights, and carrier frequency. Heights are linearly normalized, whereas frequency is logarithmically normalized to $[-1,1]$.

The VAE is pretrained from scratch using CGMs from all four carrier frequencies, with data augmentation based on rotations by multiples of 90 degrees and reflections, and is then frozen. Its shared latent representation does not explicitly take frequency or Tx/Rx heights as inputs. The training settings include Kullback--Leibler (KL) divergence regularization and mean squared error (MSE) objectives. Training hyperparameters appear in Table~\ref{tab:training_hyperparameters}.

\begin{table}[!t]
\centering
\caption{Training hyperparameters for CGM super-resolution and completion.}
\label{tab:training_hyperparameters}
\footnotesize
\setlength{\tabcolsep}{3pt}
\renewcommand{\arraystretch}{0.95}

\begin{tabular*}{\columnwidth}{
@{\extracolsep{\fill}}lr@{}}
\toprule
Parameter & Setting \\
\midrule
\multicolumn{2}{@{}l}{\textit{Shared Main-Model Training}} \\
\midrule
VAE epochs & 20 \\
VAE optimizer & AdamW \\
VAE learning rate & $10^{-4}$ \\
VAE weight decay & $10^{-4}$ \\
KL divergence weight & $10^{-6}$ \\
VAE batch size & 8 volumes \\
Selected slices per VAE volume & 4 \\
Static-model learning rate & $10^{-5}$ \\
Dynamic static-branch learning rate & $10^{-6}$ \\
Temporal/fusion-module learning rate & $10^{-6}$ \\
Frequency-conditioning learning rate & $5\times10^{-6}$ \\
Weight decay & $10^{-4}$ \\
Diffusion-time range & $[0.01,1]$ \\
$\lambda_{\mathrm{clean}}$ & 0.1 \\
$\lambda_{\mathrm{pix}}$ & 1.0 \\
$\lambda_{\mathrm{pred}}$ & 0.03 \\
$\lambda_{\mathrm{grad}}$ & 0.2 \\
$\lambda_{\mathrm{ms}}$ & 0.1 \\
$\lambda_{\mathrm{ssim}}$ & 0.2 \\
\midrule
\multicolumn{2}{@{}l}{\textit{Completion-Specific Refinement}} \\
\midrule
Refiner feature channels & 32 \\
Residual scaling factor & 0.25 \\
Refiner initial learning rate & $10^{-5}$ \\
Refiner missing-region $\ell_1$ weight & 1.0 \\
Refiner SSIM weight & 0.4 \\
Refiner gradient weight & 0.1 \\
Refiner fine-tuning learning rate & $3\times10^{-6}$ \\
Refiner fine-tuning SSIM weight & 0.25 \\
Refiner MSE weight & 1.0 \\
Refiner NMSE weight & 2.0 \\
Refiner boundary-band width & 9 pixels \\
\bottomrule
\end{tabular*}
\end{table}

The static reconstruction model is first optimized using AdamW, and the dynamic model is initialized from the best static checkpoint. During dynamic fine-tuning, the static reconstruction branch and temporal/fusion modules use a learning rate of $10^{-6}$, while the frequency-conditioning parameters use $5\times10^{-6}$. Mixed-precision training is adopted throughout. At sequence startup, the temporal history buffer and spatial history weight are initialized to zero, while subsequent frames use only previous model reconstructions. Static and dynamic checkpoints are selected based on validation reconstruction performance. For each task, the proposed ST-CGMDiff is jointly trained on the four carrier frequencies and evaluated using a single checkpoint across these frequencies. Super-resolution and completion adopt the same main-model training hyperparameters. For completion, a task-specific full-resolution refiner is additionally trained on cached 10-step coarse reconstructions and subsequently fine-tuned with additional MSE/NMSE objectives. All training hyperparameters are summarized in Table~\ref{tab:training_hyperparameters}.

\begin{figure*}[!t]
\centering
\setlength{\tabcolsep}{1pt}
\renewcommand{\arraystretch}{0.94}
\begin{tabular}{@{}cccccc@{}}

\multicolumn{3}{c}{\footnotesize $f=2.4\,\mathrm{GHz}$} &
\multicolumn{3}{c}{\footnotesize $f=3.5\,\mathrm{GHz}$}\\[-1pt]

\footnotesize $z_{\mathrm{rx}}=1.00\,\mathrm{m}$ &
\footnotesize $z_{\mathrm{rx}}=8.89\,\mathrm{m}$ &
\footnotesize $z_{\mathrm{rx}}=32.96\,\mathrm{m}$ &
\footnotesize $z_{\mathrm{rx}}=1.00\,\mathrm{m}$ &
\footnotesize $z_{\mathrm{rx}}=8.89\,\mathrm{m}$ &
\footnotesize $z_{\mathrm{rx}}=32.96\,\mathrm{m}$\\[-1pt]

\includegraphics[width=0.158\textwidth]
{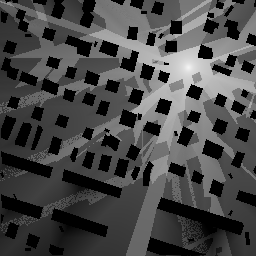} &
\includegraphics[width=0.158\textwidth]
{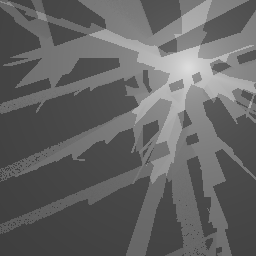} &
\includegraphics[width=0.158\textwidth]
{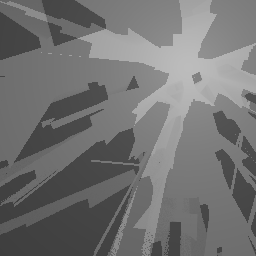} &
\includegraphics[width=0.158\textwidth]
{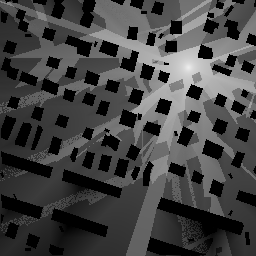} &
\includegraphics[width=0.158\textwidth]
{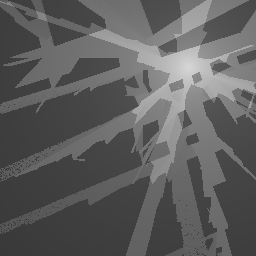} &
\includegraphics[width=0.158\textwidth]
{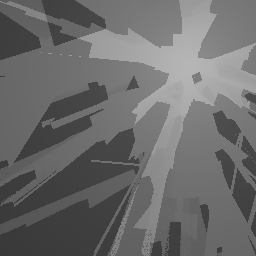}\\[1pt]

\footnotesize $t=0$ &
\footnotesize $t=7$ &
\footnotesize $t=14$ &
\footnotesize $t=0$ &
\footnotesize $t=7$ &
\footnotesize $t=14$\\[-1pt]

\includegraphics[width=0.158\textwidth]
{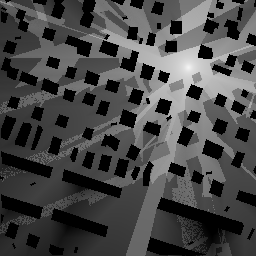} &
\includegraphics[width=0.158\textwidth]
{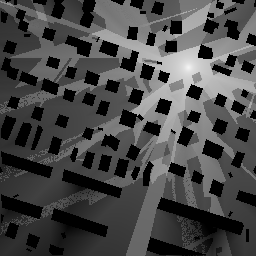} &
\includegraphics[width=0.158\textwidth]
{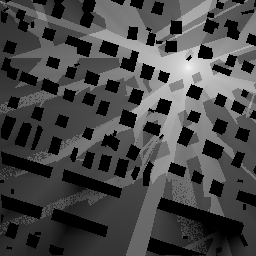} &
\includegraphics[width=0.158\textwidth]
{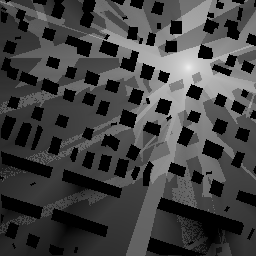} &
\includegraphics[width=0.158\textwidth]
{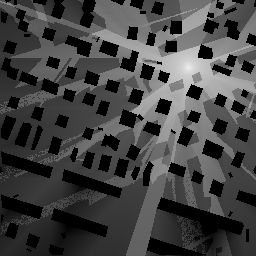} &
\includegraphics[width=0.158\textwidth]
{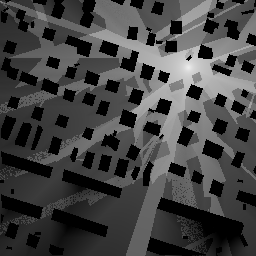}\\[3pt]

\multicolumn{3}{c}{\footnotesize $f=5.9\,\mathrm{GHz}$} &
\multicolumn{3}{c}{\footnotesize $f=28\,\mathrm{GHz}$}\\[-1pt]

\footnotesize $z_{\mathrm{rx}}=1.00\,\mathrm{m}$ &
\footnotesize $z_{\mathrm{rx}}=8.89\,\mathrm{m}$ &
\footnotesize $z_{\mathrm{rx}}=32.96\,\mathrm{m}$ &
\footnotesize $z_{\mathrm{rx}}=1.00\,\mathrm{m}$ &
\footnotesize $z_{\mathrm{rx}}=8.89\,\mathrm{m}$ &
\footnotesize $z_{\mathrm{rx}}=32.96\,\mathrm{m}$\\[-1pt]

\includegraphics[width=0.158\textwidth]
{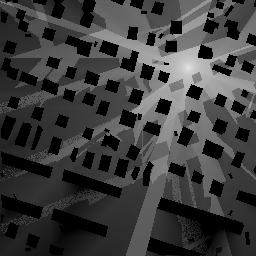} &
\includegraphics[width=0.158\textwidth]
{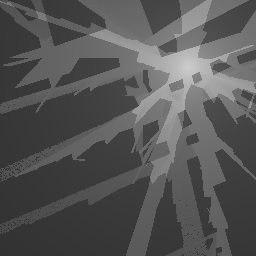} &
\includegraphics[width=0.158\textwidth]
{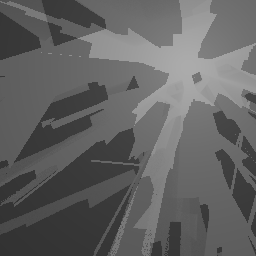} &
\includegraphics[width=0.158\textwidth]
{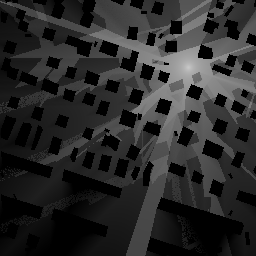} &
\includegraphics[width=0.158\textwidth]
{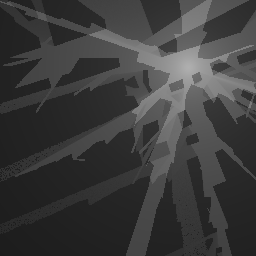} &
\includegraphics[width=0.158\textwidth]
{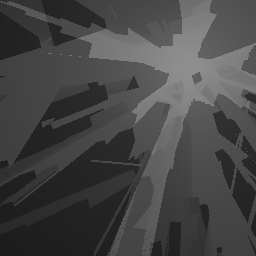}\\[1pt]

\footnotesize $t=0$ &
\footnotesize $t=7$ &
\footnotesize $t=14$ &
\footnotesize $t=0$ &
\footnotesize $t=7$ &
\footnotesize $t=14$\\[-1pt]

\includegraphics[width=0.158\textwidth]
{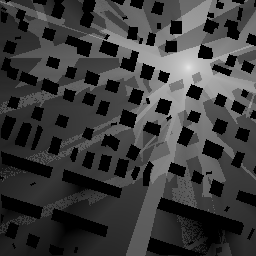} &
\includegraphics[width=0.158\textwidth]
{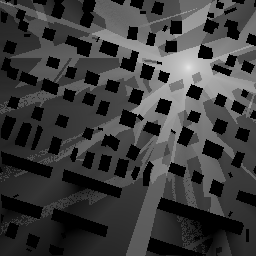} &
\includegraphics[width=0.158\textwidth]
{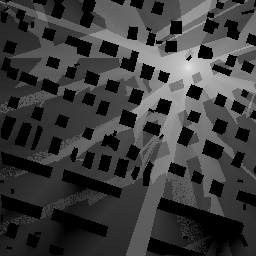} &
\includegraphics[width=0.158\textwidth]
{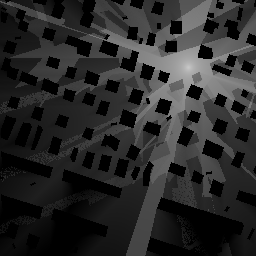} &
\includegraphics[width=0.158\textwidth]
{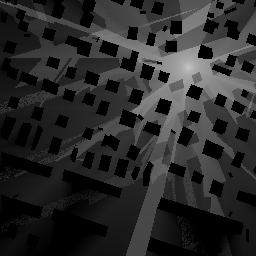} &
\includegraphics[width=0.158\textwidth]
{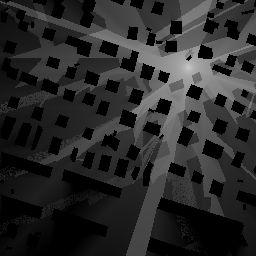}

\end{tabular}

\caption{Representative CGMs reconstructed by proposed ST-CGMDiff from the same dynamic scene at 2.4, 3.5, 5.9, and 28 GHz. For each carrier frequency, the first row shows CGMs at Rx \(z\)-coordinates of 1.00, 8.89, and $32.96\,\mathrm{m}$, while the second row shows their temporal evolution at a fixed Rx \(z\)-coordinate of $1.00\,\mathrm{m}$ and time slots $t=0$, 7, and 14. All CGMs are visualized using the same channel-gain range of $[-190,-70]\,\mathrm{dB}$.}
\label{fig:new_dataset_examples}
\end{figure*}

For comparison, we consider two diffusion-based benchmarks, RadioDiff-3D~\cite{b21} and 3D-RadioDiff~\cite{b19}, together with three conventional methods: Gaussian kernel regression~\cite{b6}, local ordinary kriging~\cite{b27}, and matrix completion~\cite{b27,b28}. RadioDiff-3D performs volumetric radio-map reconstruction using a 3D diffusion network, whereas 3D-RadioDiff reconstructs individual Rx \(z\)-layers conditioned on sparse observations and environmental information. The conventional methods recover unobserved channel values using spatial smoothing, geostatistical interpolation, or low-rank structure, respectively. Their task-specific evaluation settings are described in the corresponding comparison subsections. Unlike the proposed ST-CGMDiff, which is jointly trained, the diffusion-based baselines are trained independently at each carrier frequency and evaluated using frequency-specific checkpoints. The conventional methods do not require model training.

The experiments are conducted on NVIDIA GeForce RTX 4090 graphics processing units (GPUs). Inference latency is measured with a batch size of one.

\subsection{Cross-Frequency CGM Super-Resolution Comparison}
\label{subsec:super_resolution_comparison}

We first provide representative reconstruction results to give a qualitative impression of the capability of ST-CGMDiff. As shown in Fig.~\ref{fig:new_dataset_examples}, the proposed model reconstructs CGMs with clear spatial structures across different carrier frequencies, Rx \(z\)-coordinates, and time instants. The reconstructed maps exhibit frequency-dependent propagation patterns while preserving the variations across Rx \(z\)-layers and consecutive time slots. These qualitative results illustrate the capability of ST-CGMDiff to capture cross-frequency, 3D, and temporal channel characteristics from sparse observations.

We quantitatively compare the proposed ST-CGMDiff with the five baselines at the four training frequencies $\{2.4,3.5,5.9,28\}\,\mathrm{GHz}$. The jointly trained ST-CGMDiff is additionally evaluated at unseen $4.7$ and $15.1\,\mathrm{GHz}$ without adaptation. Each frequency has 204 sequences, evaluated at all 11 slots $t=4,\ldots,14$, totaling 2,244 test volumes.

Following its original protocol, RadioDiff-3D~\cite{b21} reconstructs the four Rx \(z\)-layers nearest to 1--$4\,\mathrm{m}$ from $10\%$ observations using 200-step denoising diffusion implicit model (DDIM) sampling. For 3D-RadioDiff~\cite{b19}, we retain its current paper-faithful evaluation protocol, where individual Rx \(z\)-layers are reconstructed from the Tx map, line-of-sight map, sparse observations, and target Rx \(z\)-coordinate, and all 273,427 selected test planes across the four training frequencies are evaluated using 1000-step denoising diffusion probabilistic model (DDPM) sampling. The conventional methods and the proposed ST-CGMDiff use random 10--50-layer subsets with layer-wise sampling rates independently drawn from $5\%$ to $10\%$. ST-CGMDiff adopts zero-start closed-loop inference with 10 reverse denoising steps. Channel gain is clipped to $[-190,-70]\,\mathrm{dB}$ and normalized to $[-1,1]$, and measured values are restored before evaluation. RMSE, NMSE, and PSNR are calculated over unobserved entries. ST-CGMDiff and the conventional methods use unobserved-entry SSIM, whereas the two diffusion baselines report SSIM over complete reconstructed outputs.

\begin{table}[!t]
\centering
\caption{Cross-frequency CGM super-resolution performance and inference latency.}
\label{tab:super_resolution_comparison}
\scriptsize
\setlength{\tabcolsep}{0.9pt}
\renewcommand{\arraystretch}{1.04}
\begin{tabular}{lcccccc}
\toprule
Method & \shortstack{Freq.\\(GHz)} &
\shortstack{Time\\(s) $\downarrow$} &
RMSE $\downarrow$ & NMSE $\downarrow$ &
SSIM $\uparrow$ & \shortstack{PSNR\\(dB) $\uparrow$}\\
\midrule
RadioDiff-3D~\cite{b21}
& Average & 0.886
& 0.19221 & 0.12572 & 0.62228 & 20.605\\
& 2.4 & --
& 0.21287 & 0.15503 & 0.59843 & 19.681\\
& 3.5 & --
& 0.20403 & 0.14280 & 0.60925 & 20.047\\
& 5.9 & --
& 0.19192 & 0.12637 & 0.61846 & 20.572\\
& 28 & --
& 0.16003 & 0.07868 & 0.66300 & 22.120\\
\midrule
3D-RadioDiff~\cite{b19}
& Average & 7.736
& 0.76363 & 6.03919 & 0.00587 & 8.389\\
& 2.4 & --
& 0.76403 & 6.12225 & 0.00500 & 8.379\\
& 3.5 & --
& 0.75719 & 6.34334 & 0.00538 & 8.458\\
& 5.9 & --
& 0.75403 & 6.67336 & 0.00560 & 8.497\\
& 28 & --
& 0.77873 & 5.04986 & 0.00745 & 8.227\\
\midrule
Gaussian kernel~\cite{b6}
& Average & \textbf{0.076}
& 0.08234 & 0.05208 & 0.83955 & 28.090\\
& 2.4 & --
& 0.09316 & 0.06894 & 0.82076 & 26.968\\
& 3.5 & --
& 0.08809 & 0.06289 & 0.82942 & 27.460\\
& 5.9 & --
& 0.08210 & 0.05356 & 0.83558 & 28.034\\
& 28 & --
& 0.06600 & 0.02291 & 0.87241 & 29.899\\
\midrule
Ordinary kriging~\cite{b27}
& Average & 0.443
& 0.09598 & 0.07039 & 0.83473 & 26.749\\
& 2.4 & --
& 0.10748 & 0.09176 & 0.81941 & 25.725\\
& 3.5 & --
& 0.10234 & 0.08476 & 0.82572 & 26.160\\
& 5.9 & --
& 0.09597 & 0.07290 & 0.83177 & 26.681\\
& 28 & --
& 0.07815 & 0.03213 & 0.86202 & 28.431\\
\midrule
Matrix completion~\cite{b27,b28}
& Average & 1.460
& 0.25388 & 0.45947 & 0.30325 & 18.072\\
& 2.4 & --
& 0.27282 & 0.55047 & 0.26338 & 17.437\\
& 3.5 & --
& 0.26408 & 0.53263 & 0.28789 & 17.714\\
& 5.9 & --
& 0.25326 & 0.48854 & 0.29381 & 18.068\\
& 28 & --
& 0.22536 & 0.26623 & 0.36791 & 19.069\\
\midrule
\textbf{ST-CGMDiff}
& Average & 0.644
& \textbf{0.06065} & \textbf{0.02915} & \textbf{0.91798} & \textbf{30.829}\\
& 2.4 & --
& \textbf{0.06923} & \textbf{0.03895} & \textbf{0.90877} & \textbf{29.621}\\
& 3.5 & --
& \textbf{0.06521} & \textbf{0.03534} & \textbf{0.91301} & \textbf{30.150}\\
& $4.7^{*}$ & --
& 0.06224 & 0.03213 & 0.91613 & 30.526\\
& 5.9 & --
& \textbf{0.06036} & \textbf{0.02965} & \textbf{0.91699} & \textbf{30.779}\\
& $15.1^{*}$ & --
& 0.05245 & 0.01915 & 0.92605 & 31.966\\
& 28 & --
& \textbf{0.04779} & \textbf{0.01264} & \textbf{0.93318} & \textbf{32.766}\\
\bottomrule
\end{tabular}
\vspace{2pt}
\par\noindent\parbox{\columnwidth}{%
\scriptsize\raggedright
\textit{Note:} Average denotes the sample-weighted mean over the four training frequencies. Frequencies marked by $^{*}$ (4.7 and 15.1 GHz) are used exclusively for testing, and only the proposed ST-CGMDiff is evaluated at these unseen frequencies without retraining or fine-tuning. SSIM is evaluated over complete reconstructed outputs for the diffusion-based baselines and over unobserved entries for other methods.\par
}
\end{table}

As shown in Table~\ref{tab:super_resolution_comparison}, the proposed ST-CGMDiff achieves the lowest average RMSE and NMSE and the highest PSNR under the evaluated protocols at the four training frequencies. Compared with RadioDiff-3D, the proposed method reduces the average RMSE and NMSE from 0.19221 and 0.12572 to 0.06065 and 0.02915, corresponding to reductions of approximately $68.4\%$ and $76.8\%$, respectively, while improving PSNR from 20.605 to $30.829\,\mathrm{dB}$. Among the conventional reconstruction methods, Gaussian kernel regression provides the strongest overall baseline performance. Compared with this baseline, the proposed method reduces RMSE and NMSE by approximately $26.3\%$ and $44.0\%$, respectively, increases SSIM from 0.83955 to 0.91798, and improves PSNR by $2.739\,\mathrm{dB}$. ST-CGMDiff achieves lower RMSE and NMSE and higher PSNR at each common frequency. The diffusion baselines retain their respective input organizations and observation budgets; these results compare the evaluated configurations.

The two unseen carrier frequencies further evaluate the interpolation generalization of the frequency-conditioned model. Without any retraining or fine-tuning, the proposed model achieves an RMSE of 0.06224, an SSIM of 0.91613, and a PSNR of $30.526\,\mathrm{dB}$ at $4.7\,\mathrm{GHz}$. At $15.1\,\mathrm{GHz}$, it achieves an RMSE of 0.05245, an SSIM of 0.92605, and a PSNR of $31.966\,\mathrm{dB}$. Notably, neither unseen frequency exhibits an obvious degradation relative to the neighboring training frequencies. These results support reconstruction at unseen frequencies within the training frequency range. They do not isolate the contribution of frequency conditioning from similarities between simulated bands.

In terms of computational cost, the proposed ST-CGMDiff requires $0.644\,\mathrm{s}$ per reconstructed volume using only 10 reverse denoising steps, compared with $0.886\,\mathrm{s}$ for the 200-step RadioDiff-3D configuration and $7.736\,\mathrm{s}$ for the current 1000-step 3D-RadioDiff timing benchmark. Gaussian kernel regression and ordinary kriging require $0.076\,\mathrm{s}$ and $0.443\,\mathrm{s}$, respectively, while matrix completion requires $1.460\,\mathrm{s}$. Because RadioDiff-3D and 3D-RadioDiff retain their original output organizations, these latency values are intended as references for computational cost rather than strictly identical-output runtime comparisons.

\subsection{CGM Completion}
\label{subsec:completion}

For completion, each selected Rx \(z\)-layer has a masked $81\times81$ square ($\approx10\%$), with the rest observed. The proposed ST-CGMDiff and 3DCKMDiff~\cite{b20} use identical masks and Rx \(z\)-layer subsets on all 2,244 samples per training frequency. The proposed method uses a jointly trained frequency-conditioned spatio-temporal model with zero-start closed-loop inference from previous reconstructions, followed by a task-specific full-resolution refiner. In contrast, 3DCKMDiff uses a separately trained checkpoint for each carrier frequency. ST-CGMDiff uses 10 reverse steps, versus 50 for 3DCKMDiff.
\begin{figure}[!t]
\centering
\setlength{\tabcolsep}{2pt}
\begin{tabular}{@{}cc@{}}
\footnotesize Incomplete CGM &
\footnotesize Completed CGM\\[-1pt]
\includegraphics[width=0.48\columnwidth]
{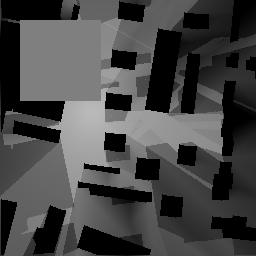} &
\includegraphics[width=0.48\columnwidth]
{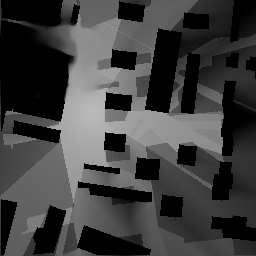}
\end{tabular}
\caption{Representative CGM completion result at $f=5.9\,\mathrm{GHz}$, time slot $t=10$, and Rx \(z\)-coordinate $z_{\mathrm{rx}}=4.20\,\mathrm{m}$. The $81\times81$ masked region in the left image denotes the unknown area, while the right image shows the CGM reconstructed by the proposed spatio-temporal completion model.}
\label{fig:completion}
\end{figure}

All metrics refer only to the unknown region: RMSE uses normalized gains, NMSE and SSIM physical gains, and PSNR the common physical gain range. The proposed ST-CGMDiff is also tested at unseen $4.7$ and $15.1\,\mathrm{GHz}$ without retraining or fine-tuning.

\begin{table}[!t]
\centering
\caption{CGM completion performance and inference latency.}
\label{tab:completion}
\scriptsize
\setlength{\tabcolsep}{1.1pt}
\renewcommand{\arraystretch}{1.05}

\begin{tabular}{lcccccc}
\toprule
Model &
\shortstack{Freq.\\(GHz)} &
\shortstack{Time\\(s) $\downarrow$} &
RMSE $\downarrow$ &
NMSE $\downarrow$ &
SSIM $\uparrow$ &
\shortstack{PSNR\\(dB) $\uparrow$}\\
\midrule

\textbf{ST-CGMDiff}
& \textbf{Average}
& \textbf{0.844}
& \textbf{0.16345}
& \textbf{0.006244}
& \textbf{0.80789}
& \textbf{23.032}\\
& 2.4 & --
& \textbf{0.18424}
& \textbf{0.008385}
& \textbf{0.79468}
& \textbf{21.910}\\
& 3.5 & --
& \textbf{0.17520}
& \textbf{0.007249}
& \textbf{0.80052}
& \textbf{22.301}\\
& $4.7^{*}$ & --
& 0.16652
& 0.006396
& 0.80647
& 22.814\\
& 5.9 & --
& \textbf{0.16306}
& \textbf{0.006000}
& \textbf{0.80749}
& \textbf{23.015}\\
& $15.1^{*}$ & --
& 0.14546
& 0.004348
& 0.81806
& 24.014\\
& 28 & --
& \textbf{0.13130}
& \textbf{0.003343}
& \textbf{0.82888}
& \textbf{24.902}\\

\midrule

3DCKMDiff~\cite{b20}
& Average
& 17.779
& 0.18407
& 0.007276
& 0.78009
& 21.552\\
& 2.4 & --
& 0.20068
& 0.009302
& 0.76518
& 20.800\\
& 3.5 & --
& 0.19456
& 0.008386
& 0.76934
& 21.010\\
& 5.9 & --
& 0.18337
& 0.007052
& 0.77747
& 21.582\\
& 28 & --
& 0.15768
& 0.004366
& 0.80838
& 22.814\\

\bottomrule
\end{tabular}
\vspace{2pt}
\par\noindent\parbox{\columnwidth}{%
\scriptsize\raggedright
\textit{Note:} Performance is evaluated over the unknown $81\times81$ region. Average denotes the mean over the four training frequencies (2.4, 3.5, 5.9, and 28 GHz). Frequencies marked by $^{*}$ (4.7 and 15.1 GHz) are used exclusively for testing, and only the proposed ST-CGMDiff is evaluated at these unseen frequencies without retraining or fine-tuning.\par
}
\end{table}
As shown in Table~\ref{tab:completion}, the completion configuration consistently outperforms the evaluated 3DCKMDiff baseline at all four training carrier frequencies and across all considered reconstruction metrics. Averaged over the 8,976 test samples from the four training frequencies, the proposed ST-CGMDiff reduces RMSE and NMSE by approximately $11.2\%$ and $14.2\%$, respectively, increases SSIM from 0.78009 to 0.80789, and improves PSNR by $1.480\,\mathrm{dB}$. The proposed method uses one jointly trained checkpoint across the four frequency bands and explicitly conditions on carrier frequency, whereas 3DCKMDiff uses a separate checkpoint for each band.

The proposed method also maintains stable completion performance at the two frequencies unseen during training. At 4.7 and $15.1\,\mathrm{GHz}$, it achieves RMSE values of 0.16652 and 0.14546 and PSNR values of 22.814 and $24.014\,\mathrm{dB}$, respectively. The resulting performance remains consistent with that of the neighboring seen frequency bands, further supporting the interpolation capability of the frequency-conditioned reconstruction framework. In addition, the proposed method reduces the number of reverse diffusion steps from 50 to 10 and decreases the average inference latency from $17.779\,\mathrm{s}$ to $0.844\,\mathrm{s}$, corresponding to a runtime reduction of approximately $95.3\%$.

\subsection{Impact of Sampling Budget}

The dynamic model trained at layer-wise $5\%$--$10\%$ rates is tested with the same checkpoint at fixed $5\%$, $7.5\%$, $10\%$, and random training-range rates. Each setting covers all 8,976 four-frequency test frames with identical Rx \(z\)-layer subsets; nested fixed-rate masks share an ordering of spatial positions. Zero-start closed-loop inference uses $K=4$ histories and 10 diffusion steps; metrics cover unobserved entries.

\begin{table}[H]
\centering
\caption{Reconstruction performance under different sampling budgets.}
\label{tab:sampling}
\footnotesize
\setlength{\tabcolsep}{4pt}
\begin{tabular}{ccccc}
\toprule
Rate & RMSE $\downarrow$ & NMSE $\downarrow$ &
SSIM $\uparrow$ & PSNR (dB) $\uparrow$\\
\midrule
$5\%$ & 0.06689 & 0.03546 & 0.90214 & 29.957\\
$7.5\%$ & 0.06028 & 0.02877 & 0.91906 & 30.870\\
$10\%$ & \textbf{0.05672} & \textbf{0.02545} & \textbf{0.92725} & \textbf{31.399}\\
Random $5\%$--$10\%$ & 0.06063 & 0.02912 & 0.91817 & 30.820\\
\bottomrule
\end{tabular}
\vspace{2pt}
\par\noindent\parbox{\columnwidth}{%
\scriptsize\raggedright
\textit{Note:} Results are evaluated on the complete four-frequency test set.\par
}
\end{table}

Increasing the fixed rate from $5\%$ to $10\%$ reduces RMSE and NMSE by $15.2\%$ and $28.2\%$, respectively, increases SSIM from 0.90214 to 0.92725, and improves PSNR by $1.441\,\mathrm{dB}$. The random-rate setting has a mean observation rate of $7.504\%$ and performs close to the fixed-$7.5\%$ setting. Its masks and Rx \(z\)-layer subsets are generated for this controlled comparison, so its results differ slightly from those in Table~\ref{tab:super_resolution_comparison}.

\subsection{Ablation Study}
\label{subsec:ablation}

We ablate the temporal ConvGRU, $z$-wise BiConvLSTM, and high-resolution refiner separately; without ConvGRU, each CGM uses only current sparse observations. All variants share 8,976 four-frequency test frames, identical Rx \(z\)-layer subsets and masks, an average of 30.15 selected layers, and a $7.504\%$ observation rate. Metrics cover unobserved entries; the full model uses the random-rate configuration in Table~\ref{tab:sampling}.

\begin{table}[t]
\centering
\caption{Ablation study of ST-CGMDiff.}
\label{tab:ablation}
\footnotesize
\setlength{\tabcolsep}{3pt}
\renewcommand{\arraystretch}{1.06}
\begin{tabular}{lcccc}
\toprule
Method & RMSE $\downarrow$ & NMSE $\downarrow$ &
SSIM $\uparrow$ & PSNR (dB) $\uparrow$\\
\midrule
\textbf{ST-CGMDiff (full)}
& \textbf{0.06063} & \textbf{0.02912} & \textbf{0.91817} & \textbf{30.820}\\
\shortstack[l]{Without Temporal\\ConvGRU}
& 0.07312 & 0.04176 & 0.88203 & 29.151\\
\shortstack[l]{Without \(z\)-wise\\BiConvLSTM}
& 0.08194 & 0.05223 & 0.81723 & 27.994\\
\shortstack[l]{Without High-Resolution\\Refiner}
& 0.10820 & 0.08974 & 0.72730 & 25.579\\
\bottomrule
\end{tabular}
\vspace{2pt}
\par\noindent\parbox{\columnwidth}{%
\scriptsize\raggedright
\textit{Note:} Results are evaluated on the complete four-frequency test set.\par
}
\end{table}

Removing any component degrades reconstruction performance. Removing the temporal ConvGRU, \(z\)-wise BiConvLSTM, and high-resolution refiner reduces PSNR by $1.669$, $2.827$, and $5.241\,\mathrm{dB}$, respectively. SSIM decreases from 0.91817 to 0.88203, 0.81723, and 0.72730. These results support the complementary contributions of temporal modeling, information exchange across Rx \(z\)-layers, and spatial refinement, with the refiner having the largest measured effect in this ablation.

\section{Conclusion}

This paper proposed ST-CGMDiff, a unified frequency-conditioned spatio-temporal latent diffusion framework for dynamic 3D CGM reconstruction from sparse measurements. By jointly exploiting temporal correlations, cross-\(z\)-layer dependencies, and frequency information, the proposed framework enables efficient reconstruction across multiple carrier frequencies and flexible Rx heights. Experiments on a dynamic multi-frequency dataset demonstrated superior reconstruction performance over the considered baselines. Compared with Gaussian kernel regression, the proposed ST-CGMDiff reduced the average RMSE by 26.3\% and improved PSNR by 2.739 dB. The framework also generalized to unseen carrier frequencies without retraining and maintained consistent performance across different sampling budgets. For contiguous-region completion, the frequency-conditioned configuration improved PSNR by 1.480 dB over 3DCKMDiff while reducing inference latency by approximately 95.3\%. Ablation results further demonstrated the complementary benefits of temporal modeling, cross-\(z\)-layer information exchange, and high-resolution refinement. In particular, the substantial performance degradation without the refiner highlights the importance of preserving fine spatial details during latent reconstruction. These findings suggest that combining spatio-temporal modeling with efficient latent diffusion provides an effective approach to dynamic 3D CGM reconstruction under limited measurement budgets.

\end{document}